\documentclass[11pt]{article}
\usepackage{graphicx} 
\usepackage{multirow}%
\usepackage{amsmath,amssymb,amsfonts}
\usepackage[margin=1.2 in]{geometry}

\usepackage{geometry,comment}
\usepackage{amsthm}%
\usepackage{mathrsfs}%
\usepackage[title]{appendix}%
\usepackage[dvipsnames,svgnames,x11names]{xcolor}
\usepackage{textcomp}%
\usepackage{manyfoot}%
\usepackage{booktabs}%
\usepackage{algorithm}%
\usepackage{algorithmicx}%
\usepackage{algpseudocode}%
\usepackage{listings}%
\usepackage{tikz}
\usetikzlibrary{arrows.meta, positioning, calc}
\usepackage{multicol}
\usepackage{cases}
\usepackage{todonotes}
\usepackage{titling}
\usepackage{graphicx} 
\usepackage[colorlinks=true,linktoc=all,linkcolor=blue,citecolor=purple,urlcolor=red,backref=page]{hyperref}
\usepackage{cleveref}
\usetikzlibrary{arrows.meta, decorations.markings}
\usepackage{graphicx} 

\date{}
\newcommand{\subtitle}[1]{%
  \posttitle{%
    \par\end{center}
    \begin{center}\LARGE#1\end{center}
    \vskip0.5em}%
}

\begin{document}
\title{\textbf{Classical Mechanics on Finite Spaces}}
\maketitle
\vspace{-3cm}
\begin{center}
{\bf Kaustav Giri\footnote{kaustavg@cmi.ac.in}, V. V. 
Sreedhar\footnote{sreedhar@cmi.ac.in}}\\
{\it Chennai Mathematical Institute, Plot H1, SIPCOT IT Park, Siruseri, 
Kelambakkam Post, Chennai 603103}\\
\vspace{1cm}
\textbf{Abstract:}
\end{center}
The connection between topology and quantum mechanics is one of the cornerstones of modern physics. Several examples of current interest like the Aharonov-Bohm effect in quantum mechanics, monopoles and instantons in quantum field theory, the quantum Hall effect in condensed matter physics, anyons in topological quantum computation, and the AdS-CFT correspondence in string theory illustrate this connection.\\

Since classical mechanics is a limiting case of quantum mechanics, it behooves us to ask how topology impacts classical mechanics. Topological considerations do play an important role in the classical context too, for example in fluid vortices and atmospheric dynamics.\\

With a desire to understand this connection more deeply, we study classical mechanics on finite spaces. Towards this end, we use the formalism developed by Bering and identify the corrections to the Klein-Gordon equation due to the presence of the boundary. We solve the modified equation in various dimensions under suitable assumptions of symmetry. 
\newpage
\section{Introduction}
A point particle will continue to be in a state of rest, or of uniform 
velocity, unless it is acted upon by an external force. A school-child 
who has gotten around to accepting this classical truth grows up to 
learn that a charged particle travelling through force-free regions can 
get scattered by a magnetic field confined to the interior of a solenoid, 
as in the case of the celebrated Aharonov-Bohm effect \cite{Aharonov1959}\cite{Ehrenberg1949}. This 
effect is variously touted as a canonical example of a quintessentially 
quantum phenomenon, as an example of the close connections between topology 
and quantum theory and, most importantly, as the authors emphasized themselves,
as proof of the physical significance of electromagnetic potentials in quantum 
theory. The claims are established either by solving the Schr\"odinger equation 
\cite{Aharonov1959}, or by incorporating the necessary modifications to the path 
integral to take into account the non-simply connected nature of the 
configuration space \cite{Gerry1979}, or by resolving the wavefunction 
into whirling waves, to which appropriate Dirac fluxes are attached before 
performing a resummation \cite{Berry1980}. In all these methods, two seemingly 
innocuous words slide into the narrative under the radar: impenetrable and 
infinitesimally-thin. The investigations in this paper stem from the 
realization that it is a mistake to  gloss over these words as trivia. In 
fact, pondering over the meaning of these words, and the role they play 
in the physics of the problem, serves to demystify the Aharonov-Bohm effect: 
neither the quantum nature of the problem, nor the physical significance of 
the vector potential are essential ingredients. The crux of the problem lies 
in impenetrability, as we argue in the following. 

Let us consider a classical, free, charged particle which we assume, without 
loss of generality, to be moving with uniform speed along the $x$-axis. Let 
a solenoid of finite cross-sectional area, carrying a steady current, be 
placed along the $z$-axis. If the solenoid were {\it not} impenetrable, the 
charged particle would enter a region with a magnetic field perpendicular to 
its velocity and would get deflected in accordance with the Newton-Lorentz 
law. On the other hand, if the solenoid is strictly impenetrable, the particle 
would bump into it and get scattered. Suppose we suffocate the deflection of the particle inside the solenoid by 
making it infinitesimally thin. The particle would still encounter 
an obstruction at the origin and get scattered, as long as the solenoid is 
impenetrable. Clearly, in this argument, it is also unimportant whether or 
not the particle is charged, and whether or not the solenoid carries a current 
and produces a magnetic field and hence a vector potential. Even a neutral, 
free particle would get scattered when it bounces off an impenetrable solenoid
carrying zero-current. How does one reconcile this logic with Newton's second law, which predicts only rectilinear motion? 

Recall that the Newton-Lorentz equation is a differential equation. Hence it
carries only local information, and is not designed to address global 
properties of the underlying space. In the absence of any forces, the 
particle would, as is its wont, travel in a straight line with a uniform 
velocity. We don't ask what it would do if, and when, it reaches the 
edge of space. We don't ask this question because we tacitly assume that 
the space is infinite in extent. To capture the effect of scattering from 
a distant obstacle, be it physical or topological, we need to formulate 
classical mechanics on finite spaces. This will require a two-fold modification
in the equations of motion: first, they can't be simple ordinary differential
equations for the position of the particle; instead they need to be formulated 
in terms of a field. Second, the usual Lagrangian and Hamiltonian formulations
of these field theories should be further modified to capture the information
about the boundary of the space. For instance, we can't blithely assume that 
the fields vanish at the boundary when we extremise the Action, and it 
will turn out that the usual definition of the Poisson bracket fails to 
satisfy the Jacobi identity \cite{Bering:1998fm}\cite{Soloviev:1993uxk}.  

We conclude this preview by mentioning that the rest of this paper is 
organised as follows. In Section 2, we review the formalism developed 
by Bering to study classical mechanics on spaces with a boundary. In 
Section 3, we will adapt Bering's formalism to the study of a massive
Klein-Gordon field in a bounded domain. In Section 4, we solve for 
the Klein-Gordon field in bounded domains of various dimensions. In Section 5, we summarize the results.\\

\section{Bering's Approach to Classical Mechanics on Finite Spaces}
A general formalism for studying Hamiltonian systems, more specifically 
classical field theories, on spaces with a boundary was developed by Bering 
\cite{Bering:1998fm}. We begin by providing a lightning review of this formalism.  
The basic idea for this formalism stems from the observation that the usual 
definition of an equal-time Poisson bracket
\begin{equation}
\label{1}
\{F(t),G(t)\}~=~\int_{\Sigma}~d^{d}x~\dfrac{\delta F(t)}{\delta\phi^{A}
(x,t)}~\omega^{AB}~\dfrac{\delta G(t)}{\delta\phi^{B}(x,t)}
\end{equation}
fails to satisfy the Jacobi identity 
\begin{equation}
\sum_{cycl.~~F,G,H}~~\{F(t),\{G(t),H(t)\}\}~=~0
\end{equation}
when the space $\Sigma$ has a boundary, $\partial \Sigma \neq~0$. Here 
$\phi^{A}(x,t)$ are classical fields. $A=1,2,3,\cdots$ labels the 
number of fields involved, for example, for a real Klein-Gordon field, $A=1,2$,  
since there are two fields {\it viz.}  $\phi^{1}=\phi$ and $\phi^{2}=\pi$, 
the momentum conjugate to $\phi$. The antisymmetric $\omega^{AB}$ is the 
usual symplectic form. The field-dependence of the functionals 
$F(t)=F\Big(\phi^{A}(x,t) \Big)$ and $G(t)=G\Big(\phi^{A}(x,t)\Big)$ 
has been suppressed to avoid clutter.
 
The variation of the functional $F(t)$ is written as follows:
\begin{equation}
\label{4}
\delta F~=~\int_{\Sigma}~d^{d}x~\sum_{k=0}^{\infty}~\partial^{k} 
\Bigg[\dfrac{\delta F}{\delta \phi^{A(k)}(x,t)}\delta\phi^{A}
(x,t)\Bigg]=\int_{\Sigma}~d^{d}x~\sum_{k=0}^{\infty}\dfrac{\partial F}
{\partial \phi^{A(k)}(x,t)}\partial^{k}\delta\phi^{A}(x,t)
\end{equation}
for arbitrary variation of field $\phi^{A}(x,t)\rightarrow 
\phi^{A}(x,t)+\delta\phi^{A}(x,t)$. Here, $\dfrac{\delta F}
{\delta \phi^{A(k)}(x,t)}$ is the higher order functional derivative, 
with $k = 0, 1,2,\cdots $ in a $d$-dimensional space. Similarly, 
$\dfrac{\partial F}{\partial \phi^{A(k)}(x,t)}$ is the higher 
order partial derivative\footnote{The higher partial derivatives of a 
functional should not be confused with the usual higher partial derivatives of
a function; they are defined by equation \eqref{4}. Both the higher order functional
derivative and the higher order partial derivative reduce to the usual 
functional derivative and the usual partial derivative when $k=0$.}. 
They are related by the following formula\footnote{In this section time `$t$' will be suppressed in $\phi^{A}(x)$ henceforth.}, 
\begin{equation}
\label{4}
\dfrac{\delta F}{\delta \phi^{A(k)}(x)}=\sum_{m \ge k} \binom{m}{k}
(-\partial)^{m-k}\dfrac{\partial F}{\partial \phi ^{A(m)}(x)}
\end{equation}
Using this general version of the functional derivative which, in particular, 
takes the surface terms into account, we define the Bering bracket of two 
arbitrary functionals $F(t)$ and $G(t)$ as,
\begin{equation}
\{F,G\}_{B}=\{F,G\}+B(F,G)\label{5}
\end{equation}
In this generalised Poisson bracket the subscript $B$ stands for Bering, $\{F,G\}$ is the usual Poisson 
bracket that comes from the bulk and $B(F,G)$ is the boundary contribution. 
$B(F,G)$ is given by 
\begin{equation}
\label{6}
B(F,G)=~\sum_{k\neq0}~\int_{\Sigma}~d^dx~ \partial ^k\Bigg[\dfrac{\delta F}
{\delta \phi^{A(k)}(x)}\omega^{AB}\dfrac{\delta G}{\delta \phi^{B}(x)}\Bigg]
-\Bigg(F \longleftrightarrow G\Bigg)
\end{equation}
and it becomes zero when our spatial boundary is at infinity.
The Poisson bracket contains partial derivatives of functionals and 
the integration in \eqref{6} is supposed to be performed inside the given bounded 
region $\Sigma$.\\

Both partial derivatives and infinitesimal volume elements
at a point require a local neighborhood of the point to be well-defined, and
may become ill-defined when that point is on the boundary. In order to circumvent this
problem, we define a suitable regularized characteristic function, 
$\chi_{\epsilon}(x)$, which is a smoothened version of the Heaviside step 
function. Using $\chi_{\epsilon}(x)$ we can extend our integration domain, 
$\Sigma$, to the entire space.
\begin{equation}
\int_{\sum} d^d x f(x)=\lim_{\epsilon\rightarrow0}\int_{\hbox{all~space}}\chi_{\epsilon}(x)
~ f_{\epsilon}(x) d^d x
\end{equation}
for any suitable test function $f(x)$ and its regularized version $f_{\epsilon}(x)$. Here, $f_{\epsilon}(x)$ is identical with $f(x)$ inside the domain $\Sigma$. For example, such a regularized characteristic function $\chi_{\epsilon}(x)$ can be taken to be
\begin{equation}
\label{8}
\chi_{\epsilon}(x)=\dfrac{1}{\left(1+\exp{\Big[-\dfrac{d_{\Sigma}(x)}{\epsilon}}
\Big]\right)}
\end{equation}
Here, $d_{\Sigma}(x)$ a metric function, is the shortest distance from the 
bulk point $x$ to the boundary. $d_{\Sigma}(x)>~0$ when $x$ is inside the bulk 
and 
$d_{\Sigma}(x)<~0$, when $x$ is outside $\Sigma$. When, $\epsilon\rightarrow0$,
$\chi_{\epsilon}(x)$ becomes $1_{\Sigma}(x)$, where,
\begin{align}
    1_{\Sigma}(x) &=1~~~if ~~x \in \Sigma\\ \nonumber
    &=0~~~{\hbox {otherwise}}
\end{align}
If $\Sigma = [a,b]$, for example, the function 
$(x-a)(b-x)$ is negative for 
$x<a$ and $x>b$, but positive for $a<x<b$. 
\subsection{Local Theories} The Hamiltonian for a local theory can be expressed as an integral over some density.\\
In this case, the functionals can be expressed as:
\begin{equation}
F[\phi^{A}]=\int_{\Sigma} d^d x f(\partial^k \phi^{A}(x),x)
\end{equation}
The functional derivative $\dfrac{\delta F}{\delta \phi^{A(k)}(x)}$ is given by,
\begin{align}
\label{11}
\dfrac{\delta F}{\delta \phi^{A(k)}(x)}&=\sum_{m \ge k} 
\binom{m}{k}(-\partial)^{m-k}\dfrac{\partial f(x)}{\partial \phi^{A(m)}(x)}\\
&=\sum_{m \ge k} \binom{m}{k}(-\partial)^{m-k}P_{A(m)}f(x)\\
&=E_{A(k)}f(x)\label{13}
\end{align}
Here, we have used the compact notation for the higher-order partial derivative $P_{A(m)}f$, and the higher Euler operator $E_{A(k)}$. Their relation can be readily obtained form \eqref{11}-\eqref{13}.\\

Since we are dealing with local theories we can put  $\dfrac{\partial f(x)}{\partial \phi^{A(m)}(x)}$ in place of $\dfrac{\partial F}{\partial \phi^{A(m)}(x)}$ unlike in  \eqref{4}. The validity of this step can be illustrated by the following simple example\cite{Greiner1996}: Let
\\
\begin{equation}
    F[\phi]=\int~(\phi(x))^n~dx
\end{equation}
Then,
\begin{align}
    \dfrac{\delta F}{\delta \phi(y)}&=\lim_{\epsilon \rightarrow0}\dfrac{1}{\epsilon}\Big(\int dx~(\phi(x)+\epsilon~\delta(x-y))^n-\int dx~ (\phi(x))^n\Big)\\
    &=\int dx ~n~(\phi(x))^{n-1}~\delta(x-y)\\
    &=n~(\phi(y))^{n-1}\\
    &=\dfrac{\partial f(y)}{\partial \phi(y)}
\end{align}
Eqaution \eqref{11} is merely a higher order version of this example.
Finally, the Bering bracket between two functionals $F$ and $G$ is 
\begin{align}
\label{Poisson Bracket}
\{F(z)&,G(w)\}_{B}=\int_{\sum \times \sum} \rho(x)d^d x \rho(y)d^d y 
\dfrac{\partial F(z)}{\partial \phi^{A(M)}(x)} \omega^{A(M)B(N)}(x,y) 
\dfrac{\partial G(w)}{\partial \phi ^{B(N)}(y)}\\ \nonumber
&=\lim_{\epsilon \to 0}\int_{\hbox{all~space}}\rho(x)\chi_{\epsilon}(x) d^d x~ \rho(y)\chi_{\epsilon}(y) 
d^d y ~\dfrac{\partial F(z)}{\partial \phi^{A(M)}(x)} \omega^{A(M)B(N)}(x,y) 
\dfrac{\partial G(w)}{\partial \phi ^{B(N)}(y)}
\end{align}
where,
\begin{align}
\label{20}
\omega^{A(M)B(N)}(x,y) =\omega ^{AB} \Big[-(-D^{\dagger}_{(x)})^{M}
(-D^{\dagger}_{(y)})^{N}+(-D^{\dagger}_{(x)})^{M} D_{(y)}^{N} +
D_{(x)}^{M}(-D^{\dagger}_{(y)})^N\Big]\delta_{\Sigma}(x,y)
\end{align}
This expression for the Bering bracket may be contrasted with the definition given in \eqref{5} and \eqref{6}. The two terms that appear in \eqref{5}, one belonging to the bulk and the second belonging to the boundary, the latter involving higher order partial derivatives, can be combined into a single term as above by replacing the partial derivative `$\partial$' with the covariant derivative `$D$' and defining a generalized symplectic kernel $\omega^{A(M)B(N)}(x,y)$ given in \eqref{20}. Note that, the measure is now $\rho(x) d^{d}x$ with volume density $\rho(x)$, and the regularized delta function $\delta_{\Sigma}(x,y)$ is given by,
\begin{equation}
    \delta_{\Sigma}(x,y)=\dfrac{\delta(x-y)}{\chi_{\epsilon}(x)\rho(x)}
\end{equation}
The advantage of using $D$ is that we do not have to worry about the factors $\binom{m}{k}$ in the derivation of  the equation of motion, to which we turn our attention presently. Although we are working in Euclidean space, it is useful to define the covariant derivative and its properties in a more abstract setting, as summarized in Appendix \ref{7.3}. We refer the reader to Bering's original paper for details \cite{Bering:1998fm}. 
\\

We begin by considering the Hamiltonian for a classical field on a bounded domain
\begin{equation}
    H=\int_{\Sigma}~d^dx~\rho(x) \mathscr{H}(x)
\end{equation}
The equation of motion is
\begin{equation}
 \dfrac{d}{dt} \phi^{A}=\{\phi^{A},H\}_{B}
\end{equation}
Putting $F(z)=\phi^{A}(z)$ and $G(w)=H$ in \eqref{Poisson Bracket}, and using
\begin{align}
\dfrac{\partial F(z)}{\partial \phi ^{A(M)}(x)}=\delta_{M,0} 
\delta_{\Sigma}(z,x)~;~~\dfrac{\partial G(w)}{\partial \phi ^{B(N)}(y)}=
\dfrac{\partial \mathscr{H}(y)}{\partial \phi^{B(N)}(y)}
\end{align}
we get
\begin{align}
\label{Proof of eqnm}
&\{\phi^{A}(z),H\}_{B} \\
&=\lim_{\epsilon \to 0}\int_{\hbox{all}}\rho(x)\chi_{\epsilon}(x) d^d x~ \rho(y)
\chi_{\epsilon}(y) d^d y ~\delta_{M,0}\delta_{\Sigma}(z,x)\omega^{A(M)B(N)}(x,y) 
\dfrac{\partial \mathscr{H}(y)}{\partial \phi ^{B(N)}(y)}\\
&=\lim_{\epsilon \to 0}\int_{\hbox{all}}\rho(x)\chi_{\epsilon}(x) d^d x~ \rho(y)\chi_{\epsilon}(y) 
d^d y ~\delta_{\Sigma}(z,x)\omega^{A(M=0)B(N)}(x,y) \dfrac{\partial 
\mathscr{H}(y)}{\partial \phi ^{B(N)}(y)}\\
&=\lim_{\epsilon \to 0}\omega^{AB} \int_{\hbox{all}}\rho(x)\chi_{\epsilon}(x) d^d x~ \rho(y)
\chi_{\epsilon}(y) d^d y ~\delta_{\Sigma}(z,x)~ \left(D^{N}_{(y)} 
\delta_{\Sigma}(x,y)\right)\dfrac{\partial\mathscr{H}(y)}{\partial 
\phi ^{B(N)}(y)}\\
&=\lim_{\epsilon \to 0}\omega^{AB} \int_{\hbox{all}} \rho(y)\chi_{\epsilon}(y) d^d y ~ \left(D^{N}_{(y)} \delta_{\Sigma}(z,y)\right)\dfrac{\partial \mathscr{H}(y)}{\partial \phi ^{B(N)}(y)}\\
&({\hbox {doing~ delta-function~integration~ on}} ~x)\\
&=\lim_{\epsilon \to 0}\omega^{AB} \int_{\hbox{all}}\rho(y) d^d y  ~\delta_{\Sigma}(z,y)(-D^{N}_{(y)})
\left(\chi_{\epsilon}(y) \dfrac{\partial \mathscr{H}(y)}{\partial \phi ^{B(N)}
(y)}\right)\\
&({\hbox{integrating~by~parts~and~$D_{(y)}^{N}~\rho(y)=0$}})\\
&=\lim_{\epsilon \to 0}\omega^{AB} \int_{\hbox{all}} \dfrac{\rho(y) \chi_{\epsilon}(y) }{\chi_{\epsilon}
(y)} d^d y  ~\delta_{\Sigma}(z,y)(-D^{N}_{(y)})\left(\chi_{\epsilon}(y) 
\dfrac{\partial \mathscr{H}(y)}{\partial \phi ^{B(N)}(y)}\right)\\
&=\lim_{\epsilon \to 0}\dfrac{\omega^{AB}}{\chi_{\epsilon}(z)}(-D^{N}_{(z)})
\left(\chi_{\epsilon}(z) \dfrac{\partial \mathscr{H}(z)}
{\partial \phi ^{B(N)}(z)}\right)\\
&=\lim_{\epsilon \to 0}\dfrac{\omega^{AB}}{\chi_{\epsilon}(z)}(-D^{N}_{(z)})\left(\dfrac{
\partial \left(\chi_{\epsilon}(z) \mathscr{H}(z)\right)}{\partial \phi ^{B(N)}
(z)}\right)\Big(\because \hbox{$\chi_{\epsilon}(z)$~is~independent~on~fields}\Big)\\
&=\lim_{\epsilon \to 0}\dfrac{\omega^{AB}}{\chi_{\epsilon}}~E_{B(0)}\left(\chi_{\epsilon}\mathscr{H}
\right)
\end{align}
Hence, the equation of motion reads:
\begin{equation}
\label{EQMN0}
    \dfrac{d}{dt} \phi^{A}=\lim_{\epsilon \to 0}\dfrac{\omega^{AB}}{\chi_{\epsilon}}~E_{B(0)}\left(\chi_{\epsilon}\mathscr{H}
\right)
\end{equation} 
This is the equation for 
a classical field on a finite space. In the next section, we 
study the specific example of the 
classical Klein-Gordon field in 
bounded domains.

\section{Klein-Gordon Field in Finite Spaces}
In this section, we formulate the equation of motion for a Klein-Gordon field in a finite space. After deriving the generalized equation of motion which encapsulates the information about the boundary, we discuss three examples with special symmetry in different dimensions, treating the corresponding boundary conditions with care. We also present a strategy for solving these equations.
Throughout the discussion, we are only concerned about spatial boundaries, no temporal boundary is considered.\\
\\
The Klein-Gordon Hamiltonian density is given by,
\begin{equation}
\label{HD}
    \mathscr{H}(\vec{r},t)=\dfrac{1}{2} {\pi}^2+\dfrac{1}{2}\Big(\vec{\nabla} \phi \Big)^2+\dfrac{1}{2}m^2\phi^2
\end{equation}
 We choose \footnote{We have used `$x$' in the previous section to denote a position vector in $\mathscr{R}^{d}$ and suppressed time `$t$' in field notation. From now on we will use $\vec{r}$ instead of `$x$' as the position vector with the usual relation in Cartesian coordinates, $\vec{r}=x\hat{i}+y\hat{j}+z\hat{k}$, along with time `$t$'. } $\phi(\vec{r},t)=\phi^{1}$ and $\pi(\vec{r},t)=\phi^{2}$ as two independent fields required to describe the field dynamics.\\
 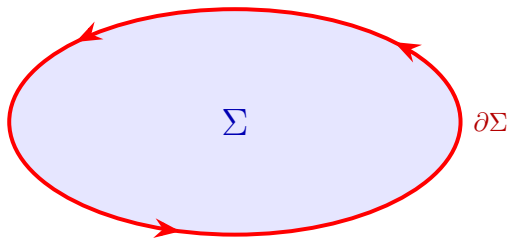
\begin{figure}[h!]
    \centering
    \begin{tikzpicture}
        \filldraw[blue!10, draw=blue!50!black, line width=0.8pt] (0,0) ellipse (3cm and 1.5cm);

        \draw[
            red,
            line width=1.5pt,
            postaction={decorate},
            decoration={
                markings,
                mark=at position 0.1 with {\arrow{Stealth}},
                mark=at position 0.4 with {\arrow{Stealth}},
                mark=at position 0.7 with {\arrow{Stealth}},
            }
        ] (0,0) ellipse (3cm and 1.5cm);

        \node[font=\Large\bfseries, text=blue!70!black] at (0,0) {$\Sigma$};
        \node[font=\small\bfseries, text=red!70!black] at (3.4, 0) {$\partial \Sigma$};
    \end{tikzpicture}
    \caption{A bounded region $\Sigma$ with a physical boundary $\partial \Sigma$.}
    \label{fig:manifold_disk_compact}
\end{figure}
\\
The equation of motion for this theory can be obtained once the higher Euler operator appearing in \eqref{EQMN0} is known. This is easily obtained from \eqref{11}-\eqref{13} 
\begin{equation}
\label{EO}
     E_{B(0)}f=\sum_{m \geq 0} \binom{m}{0} (-\partial)^{m}\Big(\dfrac{\partial f}{\partial \phi^{B(m)}}\Big)
\end{equation}
The rules governing higher Euler operators $E_{B(k)}$ are discussed in Appendix \ref{7.2}. \\

The Klein-Gordon Hamiltonian density contains only up to first-order spatial derivatives of the fields. Therefore, in \eqref{EO}, $m=1$ is the maximum value.
Using, \eqref{EQMN0},\eqref{HD} and \eqref{EO} and remembering that $\chi_{\epsilon}$ with $\epsilon \to 0$, does not depend on the fields, we get 
\begin{equation}
\label{Field}
    \dfrac{d \phi^{1}}{dt}=\dfrac{d \phi}{dt}
    =\dfrac{\omega^{12}}{\chi_{\epsilon}}E_{2(0)}\Big(\chi_{\epsilon}\mathscr{H}\Big)
\end{equation}
\begin{align}
    &=\dfrac{\omega^{12}}{\chi_{\epsilon}}
\sum_{m \geq 0} \binom{m}{0} (-\partial)^{m}\Big(\dfrac{\partial (\chi_{\epsilon}\mathscr{H})}{\partial \phi^{ 2 (m)}}\Big)=\dfrac{\omega^{12}}{\chi_{\epsilon}}
\sum_{m \geq 0}\binom{m}{0} (-\partial)^{m} \Big(\dfrac{\partial (\chi_{\epsilon}\mathscr{H})}{\partial \pi^{  (m)}}\Big)\\
&=\pi
\end{align}  
and,
\begin{align}
\label{Momentum}
    &\dfrac{d \phi^{2}}{dt}=\dfrac{d \pi}{dt}
    =\dfrac{\omega^{21}}{\chi_{\epsilon}}E_{1(0)}\Big(\chi_{\epsilon}\mathscr{H}\Big)\\
    &=\dfrac{\omega^{21}}{\chi_{\epsilon}}
\sum_{m \geq 0} \binom{m}{0} (-\partial)^{m}\Big(\dfrac{\partial (\chi_{\epsilon}\mathscr{H})}{\partial \phi^{  (m)}}\Big) \label{43}\\
    &=-\dfrac{1}{\chi_{\epsilon}}\Big[\chi_{\epsilon}\dfrac{\partial \mathscr{H}}{\partial \phi}-\dfrac{\partial}{\partial x}\Big(\dfrac{\partial(\chi_{\epsilon} \mathscr{H})}{\partial \Big(\dfrac{\partial \phi}{\partial x}\Big)}\Big)-\dfrac{\partial}{\partial y}\Big(\dfrac{\partial(\chi_{\epsilon} \mathscr{H})}{\partial \Big(\dfrac{\partial \phi}{\partial y}\Big)}\Big)-\dfrac{\partial}{\partial z}\Big(\dfrac{\partial(\chi_{\epsilon} \mathscr{H})}{\partial \Big(\dfrac{\partial \phi}{\partial z}\Big)}\Big)\Big]\\
    &=-m^2 \phi+\Big(\dfrac{\partial ^2 \phi}{\partial x^2}\Big)+\Big(\dfrac{\partial ^2 \phi}{\partial y^2}\Big)+\Big(\dfrac{\partial ^2 \phi}{\partial z^2}\Big)+\dfrac{1}{\chi_{\epsilon}}\dfrac{\partial \chi_{\epsilon}}{\partial x} \dfrac{\partial \phi}{\partial x}+\dfrac{1}{\chi_{\epsilon}}\dfrac{\partial \chi_{\epsilon}}{\partial y} \dfrac{\partial \phi}{\partial y}+\dfrac{1}{\chi_{\epsilon}}\dfrac{\partial \chi_{\epsilon}}{\partial z} \dfrac{\partial \phi}{\partial z}\label{45}
\end{align}
From \eqref{Field} and \eqref{Momentum}, it follows
\begin{equation}
\label{eqnm-0}
    \dfrac{d^2 \phi}{d t^2}-\Big(\dfrac{\partial ^2 \phi}{\partial x^2}\Big)-\Big(\dfrac{\partial ^2 \phi}{\partial y^2}\Big)-\Big(\dfrac{\partial ^2 \phi}{\partial z^2}\Big)+m^2 \phi=\lim_{\epsilon \to 0}\Big[\dfrac{1}{\chi_{\epsilon}}\dfrac{\partial \chi_{\epsilon}}{\partial x} \dfrac{\partial \phi}{\partial x}+\dfrac{1}{\chi_{\epsilon}}\dfrac{\partial \chi_{\epsilon}}{\partial y} \dfrac{\partial \phi}{\partial y}+\dfrac{1}{\chi_{\epsilon}}\dfrac{\partial \chi_{\epsilon}}{\partial z} \dfrac{\partial \phi}{\partial z}\Big]
\end{equation}
\begin{equation}
\label{EQM-0}
    \Rightarrow\dfrac{d^2 \phi}{d t^2}-\vec{\nabla}^2 \phi+m^2 \phi=\lim_{\epsilon \to 0}\dfrac{1}{\chi_{\epsilon}}\Big(\vec{\nabla}\chi_{\epsilon}\cdot\vec{\nabla} \phi\Big)
\end{equation}
\\
This is the coveted equation for the Klein-Gordon field in a finite three-dimensional space. \\
\subsection{The Bulk and the Boundary:}

As mentioned above, we will now consider simple examples in various dimensions with symmetric domains. In one dimension, $\chi_{\epsilon}$ is a 
function of $x$ and, in two and three dimensions, we choose the characteristic function $\chi_\epsilon$ to be only a function of the radial coordinate.
Hence, 
\begin{align}
    \dfrac{d^2 \phi}{d t^2}-\vec{\nabla}^2 \phi+m^2 \phi
     &=\lim_{\epsilon \to 0}\dfrac{1}{\chi_{\epsilon}}\dfrac{d \chi_{\epsilon}}{d x} \dfrac{\partial \phi}{\partial x}\label{48}~~~~~[{\hbox{in}}~1\hbox{D}]\\
     &=\lim_{\epsilon \to 0}\dfrac{1}{\chi_{\epsilon}}\dfrac{d \chi_{\epsilon}}{d r} \dfrac{\partial \phi}{\partial r}\label{49}~~~~~ [{\hbox{in}} ~2 \hbox{D}~{\hbox{and}} ~3\hbox{D}]
\end{align}
where, in \eqref{48} and \eqref{49}, 
we call $\dfrac{1}{\chi_{\epsilon}}\dfrac{d \chi_{\epsilon}}{d x}$, the `weight function'.\\ 

As already explained in \eqref{8}, the characteristic function we choose is given by
\begin{equation}
    \chi_{\epsilon}(x)=\dfrac{1}{1+\exp{\Big[-\dfrac{d_{\Sigma}(x,z)}{\epsilon}\Big]}}
\end{equation}
where $x$ and $z$ are points in interior($\Sigma$) and on the 
boundary($\partial \Sigma$) respectively. When $x$ is a point inside the domain, $d_{\Sigma}(x,z)$ is positive, and when $x$ is outside the domain, it is negative. For example, in a one-dimensional connected closed interval $[a,b]$, the $\chi_{\epsilon}(x)$ is,
\begin{equation}
    \chi_{\epsilon}(x)=\dfrac{1}{1+\exp{\Big(-\dfrac{(x-a)(b-x)}{\epsilon}\Big)}}
\end{equation}
and in two and three dimensions the rotationally symmetric $\chi_{\epsilon}(r)$ takes the following form, 
\begin{equation}
    \chi_{\epsilon}(r)=\dfrac{1}{1+\exp\Big(-\dfrac{(a-r)}{\epsilon}\Big)}
\end{equation}
\\
where $a$ is the radius of the boundary in plane polar and spherical polar coordinates respectively.

\begin{figure}[h!]
    \centering
    \begin{tabular}{cc}
        \includegraphics[width=0.3\textwidth]{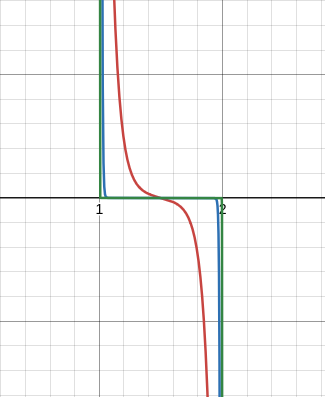} &
        \includegraphics[width=0.3\textwidth]{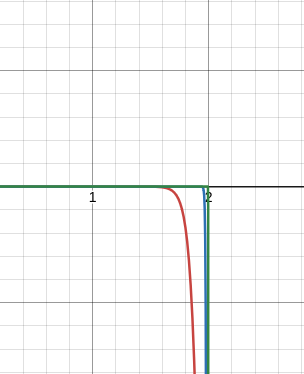} \\
        (a) 1D & (b) 2D and 3D
    \end{tabular}
    \caption{Plots of `weight functions vs position' for $\epsilon=0.05$,$\epsilon=0.005$ $\hbox{and}$ $\epsilon=0.0005$.}
    \label{Weight_Function}
\end{figure}

We observe that the weight functions $\dfrac{1}{\chi_{\epsilon}}\dfrac{d \chi_{\epsilon}}{d x}$ and $\dfrac{1}{\chi_{\epsilon}}\dfrac{d \chi_{\epsilon}}{d r}$ in \eqref{48},\eqref{49} become large when approaching the end points of the interval and near the boundary respectively as $\epsilon \rightarrow 0$. In the bulk, they become zero as $\epsilon \rightarrow 0$. It is important to note that there is always an arbitrarily small interval irrespective of how small $\epsilon$ becomes, which separates the bulk from the boundary. The boundary can only be reached asymptotically -- reminiscent of Zeno's paradox \cite{sep-paradox-zeno}. In this interval, the weight function takes a finite value. The weight functions are plotted against the appropriate coordinates in the given figures (\ref{Weight_Function}). In (a)~the plot is for a one-dimensional interval $[1,2]$ and in (b), the plot is for two and three dimensions with radius $r=2$ in appropriate units.
In both plane polar and spherical polar coordinates, $\hat{r} \cdot \vec{\nabla}=\dfrac{\partial}{\partial r}$ and the direction of the normal to the boundary is always along the radial direction for our rotationally symmetric domains, 
$\hat{n}=\hat{r}$.\\

To navigate through the twilight zone between the bulk and boundary, the weight function may be represented by a sequence of functions $g_n(\vec r \to \partial\Sigma )$ in terms of which \eqref{48} and \eqref{49} can be written as,
\begin{equation}
\label{TDE}
    \dfrac{d^2 \phi(\vec{r},t)}{d t^2}-\vec{\nabla}^2 \phi(\vec{r},t)+m^2 \phi(\vec{r},t)=-\lim_{n\rightarrow\infty}~g_{n}(\vec{r}\to\partial \Sigma)(\vec{\nabla}\phi \cdot \hat{r})
\end{equation}
As $n \rightarrow~\infty$, $g_{n}(\vec{r}\to\partial \Sigma)$ is $\infty$  when $\vec{r}$ approaches any point on the `boundary' $\partial \Sigma$, and zero elsewhere. Thus, the sequence of functions $g_n(\vec r\to\partial\Sigma)$ behaves like a delta function as $n \to \infty$, although care needs to 
be exercised in defining the {\it value} of the delta function on the boundary. The negative sign on the right hand side of \eqref{TDE} reflects the negative increment of the weight function mentioned earlier. This equation is in the twilight zone between being homogeneous and inhomogeneous, for although the right hand side is linear in $\phi$, at the boundary $\vec{\nabla}\phi\cdot \hat{r}$ can take any specified non-zero value in which case the equation is inhomogeneous and the right hand side acts like a source.  \\

We define the sequence of functions $g_{n}(x)$ in one dimension\cite{olver2014introduction} to be,
\begin{equation}
    g_{n}(x)=\dfrac{n}{\pi(1+n^2x^2)}
\end{equation}
This function has the property of being $+\infty$ at $x=0$ and becomes zero elsewhere when $n \to \infty$. Further, we have 
\begin{equation}
    \int_{-\infty}^{+\infty}~\dfrac{n}{\pi(1+n^2x^2)}~dx=~1
\end{equation}
for all $n$. Hence, 
\begin{equation}
\label{delta function_0}
    \lim_{n \to \infty}~g_{n}(\vec{r}\to\vec{r}_{0})~=~\delta(\vec{r}-\vec{r}_{0})~~\hbox{where}~\vec{r}_{0}\in \partial\Sigma.
\end{equation}
Moreover, by modifying the limits of the integral, we have the following result 
\begin{equation}
     f_{n}(x)=\int_{-\infty}^{x}~\dfrac{n}{\pi(1+n^2t^2)}~dt=\dfrac{1}{\pi}~\tan^{-1}(nx)+\dfrac{1}{2}
\end{equation}
It follows that
\begin{align}
\label{61}
    \lim_{n\rightarrow \infty}~f_{n}(x)=\sigma(x)&= 0, ~~~ x < 0\\
    &=\dfrac{1}{2}, ~~~x=0\\
    &=1,~~~x >0
\end{align}
Therefore,  
\begin{equation}
    \int_{ -\infty}^{0} \delta(x-0) ~dx=\dfrac{1}{2}
\end{equation}
The Fourier expansion of the unit~step~function $\sigma(x)$ in \eqref{61}, whose derivative is the Dirac delta function, also confirms
the assigned value $\dfrac{1}{2}$ at $x=0$.\\

Thus, we define the generalized delta function $\tilde\delta (\vec r -\vec r_0)$ as follows
\begin{align}
   \int_{\Sigma} \Tilde{\delta}(\vec{r}-\vec{r}_{0})dV_{\vec{r}}~&=~1~~~~{\hbox{when}}~\vec{r}_{0} \in \Sigma \label{Delta-1}\\
    &=\dfrac{1}{2}~~~~{\hbox{when}}~\vec{r}_{0} \in \partial \Sigma\label{Delta-2}
\end{align}

\subsection{Helmholtz Equation in Finite Spaces} Since we are interested in the effect of spatial boundaries, we 
can separate out the time variable. It is well-known that separation of variables in the Klein-Gordon equation leads to the Helmholtz equation. Since the boundary correction to the Klein-Gordon equation depends only on the gradient of $\phi(\vec{r},t)$, and not on the time-derivative, it turns out that a separation of variables can nevertheless be carried out even in this case, as shown below. Substituting the factorization,
\begin{equation}
\label{SV}
    \phi(\vec{r},t)=\psi(\vec{r})T(t)
\end{equation}
in the Klein-Gordon equation and separating variables gives \\
\begin{equation}
    \dfrac{d^2 T}{dt^2}=-\omega^2 T
\end{equation}
and 
\begin{equation}
\label{65}
   \vec{\nabla}^2 \psi(\vec{r})+(\omega^2-m^2) \psi(\vec{r})= \vec{\nabla}^2 \psi(\vec{r})+\kappa^2 \psi(\vec{r})~=~\lim_{n\rightarrow\infty}~g_{n}(\vec{r}\to\partial \Sigma)(\vec{\nabla}\psi \cdot \hat{r})
\end{equation}
As already explained, the source term on the right hand side has its support only on the spatial boundary, hence, from \eqref{Delta-1} and \eqref{Delta-2}, the above equation can be rewritten as,
\begin{equation}
\label{EQNM-1}
    \vec{\nabla}^2 \psi(\vec{r})+\kappa^2 \psi(\vec{r})~=~\int_{\partial \Sigma}\Tilde{\delta}(\vec{r}-\vec{r}_{0})~(\vec{\nabla}\psi \cdot \hat{n})\Big|_{\partial \Sigma}~dS_{0}
\end{equation}
where $\vec{r}_{0}$ are the variables on the boundary $\partial \Sigma$ and 
$\kappa^2 = \omega^2 -m^2$.\\

A formal proof of this restriction to the boundary follows. Integrating both sides of \eqref{65} after regularizing the right hand side by the characteristic function, we get, \begin{align}
\label{72}
   &\int_{\Sigma} \vec{\nabla}^2 \psi(\vec{r})~dV+\int_{\Sigma }\kappa^2 \psi(\vec{r})~dV~\\
   &=\lim_{\epsilon \rightarrow0}~\lim_{n\rightarrow\infty}\int_{\hbox{all}}~\chi_{\epsilon}(r)~g_{n}(\vec{r}\to\partial \Sigma)(\vec{\nabla}\psi \cdot \hat{n})~dV\\
    &=\lim_{\epsilon \rightarrow0}~\lim_{n\rightarrow\infty}\int_{\hbox{all}}
     \dfrac{1}{\left(1+\exp{\Big[-\dfrac{d_{\sum}(r)}{\epsilon}}\Big]\right)}~g_{n}(\vec{r}\to\partial \Sigma)(\vec{\nabla}\psi \cdot \hat{n})~dV\\
     &=\lim_{\epsilon \rightarrow0}\int_{\hbox{all}}\int_{\partial \Sigma}
     \dfrac{1}{\left(1+\exp{\Big[-\dfrac{d_{\sum}(r)}{\epsilon}}\Big]\right)}\delta_{\hbox{all}}(\vec{r}-\vec{r}_{0})(\vec{\nabla}\psi \cdot \hat{n})~dS_{0}~dV\\    
     &=\dfrac{1}{2}\int_{\partial \Sigma}(\vec{\nabla}\psi \cdot \hat{n})|_{\partial \Sigma}~dS_{0}~~~~\Big[\because d_{\Sigma}(r)\rightarrow 0,~\hbox{as}~\vec{r}\rightarrow\vec{r}_{0}\Big]\\
     &=\int_{\Sigma}\int_{\partial \Sigma}\Tilde{\delta}(\vec{r}-\vec{r}_{0})(\vec{\nabla}\psi \cdot \hat{n})\Big|_{\partial \Sigma}~dS_{0}~dV
\end{align}
\\
Here the domain of $\delta_{\hbox{all}}(\vec{r}-\vec{r}_{0})$ is the entire space $\mathcal{R}^{n}$. Equating integrands on both sides we get the equation of motion~\eqref{EQNM-1}. \\  

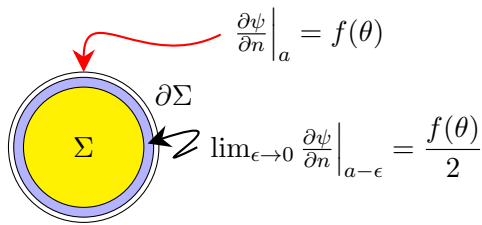
\begin{figure}[h!]
    \centering
    \begin{tikzpicture}
        \coordinate (center) at (0,0);
        \draw[fill=white!30, draw=black] (center) circle (1cm);
        \draw[fill=blue!30, draw=black] (center) circle (0.93cm);
        \draw[fill=yellow, draw=black] (center) circle (0.8cm);
        \node at (1.2, 0.7){$\partial \Sigma$};
        \node at (0.0,0.0){$\Sigma$};
        \node (derivative_text) at (3.5, 0)
            {$\lim_{\epsilon\to 0} \frac{\partial \psi}{\partial n}\Big|_{a-\epsilon} =\dfrac{f(\theta)}{2} $};
        \coordinate (strip_point) at (0.85, 0.05); 

        \draw[-{Stealth[length=3mm, width=3mm]}, thick, black] (derivative_text.west)
            .. controls ($(derivative_text.west) - (1cm, 0.5cm)$) and ($(strip_point) + (1.5cm, 0.5cm)$) .. (strip_point);
        \node (derivative_text) at (3.0, 1.5)
            {$ \frac{\partial \psi}{\partial n}\Big|_{a} =f(\theta) $};
        \coordinate (strip_point) at (0.0, 1.0);
        \draw[-{Stealth[length=3mm, width=3mm]}, thick, red] (derivative_text.west)
           .. controls ($(derivative_text.west) - (1.0cm, 0.5cm)$) and ($(strip_point) + (0.0cm, 1.0cm)$) .. (strip_point);
    \end{tikzpicture}
    \caption{Bounded space $\Sigma$ of radius $r=a$ with boundary $\partial \Sigma$ in 2D.}
    \label{fig:concentric_annulus_arrow}
\end{figure}

The function $\vec{\nabla}\psi \cdot \hat{n}$ has a $\it{jump}$ discontinuity at the boundary. For simplicity, considering \eqref{EQNM-1} in  one dimension and performing integration from $a-\epsilon$ to $a$, the point on the boundary, with $\lim_{\epsilon \to 0}$ we get,\\
\begin{align}
    &\lim_{\epsilon \to 0}\int_{a-\epsilon}^{a} \dfrac{d^2 \psi}{dx^2}~dx+\lim_{\epsilon \to 0}\int_{a-\epsilon}^{a} ~\kappa^2\psi~dx~=~\lim_{\epsilon \to 0}\int_{a-\epsilon}^{a}\Tilde{\delta}(x-a)\dfrac{d\psi}{dx}\Big|_{x=a}~dx\\
    &\Rightarrow \dfrac{d\psi}{dx}\Big|_{x=a}-\lim_{\epsilon \to 0}\dfrac{d\psi}{dx}\Big|_{x=a-\epsilon}~=~\dfrac{1}{2}\dfrac{d\psi}{dx}\Big|_{x=a}\\
    &\Rightarrow \lim_{\epsilon \to 0}\dfrac{d\psi}{dx}\Big|_{x=a-\epsilon}~=~\dfrac{1}{2}\dfrac{d\psi}{dx}\Big|_{x=a}\label{C1}
\end{align}
\\
The generalized $\it{Neumann}$ boundary condition that takes into account the $\it{jump~discontinuity}$ consists in specifying $\lim_{\epsilon \to 0}\dfrac{\partial \psi}{\partial x}\Big|_{x=a-\epsilon}$ to be some constant $\dfrac{\alpha}{2}$, say, in one dimension, and $\lim_{\vec{r}\to \partial \Sigma}(\vec{\nabla}\psi \cdot \hat{n})=\dfrac{f(\theta, \phi)}{2}$ in higher dimensions for a given function of the boundary variables $\vec{\nabla}\psi\cdot \hat{n}\Big|_{\partial \Sigma}=f(\theta, \phi)$. A pictorial representation of the generalized Neumann boundary condition in two dimensions, is shown in  (\ref{fig:concentric_annulus_arrow}).
\\

We refer to $\vec{\nabla}\psi\cdot \hat{n}\Big|_{\partial \Sigma}$ as $\it{boundary ~condition}$ and $\lim_{\vec{r}\to \partial \Sigma}(\vec{\nabla}\psi \cdot \hat{n})$ as $\it{effective~boundary}$ $\it{condition} $ respectively.
The difference between the two encodes the jump discontinuity.\\ 

We conclude this section by mentioning the crucial points that distinguish the Klein-Gordon theory on spaces with boundary from its counterpart on spaces without a boundary \eqref{EQNM-1}.
First, the form of the right hand side of \eqref{EQNM-1}, in presence of a physical boundary, suggests that we use Neumann boundary condition. Unlike the 
Poisson equation, the resulting source term has support only on the boundary.
Second, it is useful to compare the Klein-Gordon equation with Neumann boundary conditions, with and without the boundary terms. The equation in the bulk is exactly the same in both cases, since the boundary delta function vanishes in the bulk. If the boundary correction is not taken into account, we have a homogeneous equation. On the other hand, if it is taken into account, the equation starts to become inhomogeneous as we approach the boundary, exactly on which the source term becomes infinite, reflecting the 
impenetrability of the boundary. 
Consequently, the solution of the Klein-Gordon equation with the boundary terms should encode the information of impenetrability.
\\
\\
Finally, we have, for the sake of simplicity, restricted our attention to spaces which enjoy rotational symmetry. 
Hence, the normal to the boundary $\hat{n}$ is the same as radial unit vector $\hat{r}$. This will come in 
handy when we illustrate the results in the special examples worked out in the next section. For a domain with an arbitrary shape, the normal direction has all three components $\hat{r},\hat{\theta}, \hat{ \phi}$. In this 
case we have to choose  $\vec{\nabla}\chi_{\epsilon}$ in \eqref{EQM-0} such that, $\dfrac{\vec{\nabla}\chi_{\epsilon}}{|\vec{\nabla}\chi_{\epsilon}|}=\hat{n}$. The equation of motion of the Klein-Gordon field \eqref{EQNM-1}, however is valid for $\it{any}$ finite domain.
\\
\section{Solutions of Klein-Gordon Equation in Finite Spaces}\label{section 3.4}
In \eqref{SV} we applied the separation of variable method to derive the time-independent equation of motion \eqref{EQNM-1}. It is clear that we can solve this equation of motion once the boundary condition is specified in the form of $(\vec{\nabla}\psi \cdot \hat{n})\Big|_{\partial \Sigma}$. The formal way of assigning a boundary condition to a dynamical field is the following, 
\begin{equation}
    \vec{\nabla}\phi(\vec{r},t)\cdot \hat{n}\Big|_{\partial \Sigma}=g(t,\alpha_{i})
\end{equation}
here, $g(t,\alpha_{i})$ is a function of time and boundary coordinates, $\alpha_i$.\\
Unfortunately, for a general function $g(t, \alpha_{i})$, the `separation of variable' method is not always suitable. Therefore, we restrict our attention to the boundary condition of the following form,
\begin{equation}
    \vec{\nabla}\phi(\vec{r},t)\cdot \hat{n}\Big|_{\partial \Sigma}=g(t,\alpha_{i})=f(\alpha_{i})T(t)
\end{equation}
\begin{equation}
    \Rightarrow T(t)~\Big(\vec{\nabla}\psi(\vec{r})\cdot \hat{n}\Big|_{\partial \Sigma}\Big)=f(\alpha_{i})T(t)
\end{equation}
\begin{equation}
    \Rightarrow \Big(\vec{\nabla}\psi(\vec{r})\cdot \hat{n}\Big)\Big|_{\partial \Sigma}=f(\alpha_{i})
\end{equation}
\\
Next, we write down the solution $\psi(\vec{r})$ of \eqref{EQNM-1} as a sum of a homogeneous solution and  the particular integral.
\begin{equation}
\label{80}
    \psi(\vec{r})=\Phi_{0}(\vec{r})+\Phi_{1}(\vec{r})
\end{equation}
Here, $\Phi_{0}(\vec{r})$ is the homogeneous solution and $\Phi_{1}(\vec{r})$ is the particular integral. Thus,
\begin{align}
     &(\vec{\nabla}^2+\kappa^2)~\Phi_{0}(\Vec{r})=0\label{HOM}\\
     &(\vec{\nabla}^2+\kappa^2)~\Phi_{1}(\Vec{r})~=~\int_{\partial \Sigma} \Tilde{\delta}(\vec{r}-\vec{r}_{0})~(\vec{\nabla}\psi \cdot \hat{n})\Big|_{\partial \Sigma}~dS_{0}\label{PI}
 \end{align}
The particular integral $\Phi_{1}(\vec{r})$, can be written as the linear superposition of the 
corresponding Green's function $G(\vec{r},\vec{r}_{0})$ of the Helmholtz operator, multiplied by given boundary condition as weight factors.
\begin{equation}
\label{DI}
    \Phi_{1}(\vec{r})~=~\int_{\partial \Sigma}~G(\vec{r},\vec{r}_{0})(\vec{\nabla}\psi \cdot \hat{n})|_{\partial \Sigma}~dS_{0}
\end{equation}
Here, 
\begin{equation}
    (\vec{\nabla}^2+\kappa^2)G(\vec{r}, \vec{r}_{0})=\Tilde{\delta}(\vec{r}-\vec{r}_{0})\label{G0}
\end{equation}
Both $\psi(\vec{r})$ and $\Phi_{1}(\vec{r})$ have a $\it{jump}$ discontinuity at the boundary. But, being a solution of the homogeneous equation, $\Phi_{0}(\vec{r})$ is continuous on the boundary. Therefore,
\begin{equation}
    \lim_{\vec{r}\to \partial \Sigma}~\Big(\vec{\nabla}\Phi_{0}(\vec{r})\cdot \hat{n}\Big)~=~\vec{\nabla}\Phi_{0}(\vec{r})\cdot \hat{n}\Big|_{\partial \Sigma}\label{HBC}
\end{equation}
From the linearity of the Helmholtz differential operator and taking into account the jump discontinuity\eqref{C1} we get, 
\begin{equation}
    (\vec{\nabla}\psi \cdot \hat{n})\Big|_{\partial \Sigma}~=~
    (\vec{\nabla}\Phi_{0} \cdot \hat{n})\Big|_{\partial \Sigma}+ (\vec{\nabla}\Phi_{1} \cdot \hat{n})\Big|_{\partial \Sigma}\label{LIM}
\end{equation}
and,
\begin{equation}
    \vec{\nabla}G(\vec{r},\vec{r}_{0})\cdot \hat{n}\Big|_{\vec{r}_{0}\in\partial \Sigma}- \lim_{\vec{r}\to \vec{r}_{0}\in \partial \Sigma}\Big(\vec{\nabla}G(\vec{r},\vec{r}_{0})\cdot \hat{n}\Big)~=~\dfrac{1}{2}\label{G1}
\end{equation}
\\
We will calculate the Green's function in \eqref{G0} in different dimensions using the method of eigenfunction expansions. We choose the Green's function to satisfy the $\it{homogeneous}$ Neumann boundary condition. e.g., 
$\Big(\vec{\nabla}G(\vec{r},\vec{r}_{0})\cdot \hat{n}\Big)\Big|_{\partial \Sigma}=0$.
From \eqref{DI} it follows that 
\begin{equation}
    (\vec{\nabla}\Phi_{1} \cdot \hat{n})\Big|_{\partial \Sigma}=0\label{BC1}
\end{equation}
Hence from \eqref{LIM} and \eqref{HBC}, 
\begin{equation}
    \lim_{\vec{r}\to \partial \Sigma}(\vec{\nabla}\Phi_{0} \cdot \hat{n})~=~(\vec{\nabla}\Phi_{0} \cdot \hat{n})\Big|_{\partial \Sigma}=(\vec{\nabla}\psi \cdot \hat{n})\Big|_{\partial \Sigma}=f(\theta, \phi)\label{BC2}
\end{equation}
Therefore, \eqref{BC2} and \eqref{BC1} are the boundary conditions followed by the homogeneous solution and the particular integral of the total solution respectively, provided $(\vec{\nabla}\psi \cdot \hat{n})|_{\partial \Sigma}=f(\theta,\phi)$.\\

Since, $\Phi_{1}$ will be written in terms of sum of eigenfunctions of helmholtz operator following the $\it{homogeneous}$ Neumann boundary condition and it's derivative has a $\it{jump}$ discontinuity at the boundary. From \eqref{G1}
\begin{align}
    \lim_{\vec{r}\to \vec{r}'_{0}\in \partial \Sigma}\vec{\nabla}\Phi_{1}\cdot\hat{n}&=\lim_{\vec{r}\to \vec{r}'_{0}\in\partial \Sigma}\int_{\partial \Sigma}\Big(\vec{\nabla}G(\vec{r},\vec{r}_{0})\cdot \hat{n}\Big)(\vec{\nabla}\psi \cdot \hat{n})|_{\partial \Sigma}~dS_{0}\\
    &=-\dfrac{1}{2}(\vec{\nabla}\psi \cdot \hat{n})|_{\vec{r}'_{0}}\label{limpoi}
\end{align}
In the last line, we have used the fact that, when $\vec{r}\to\vec{r}'_{0}\neq\vec{r}_{0}$,
\\
\begin{equation}
    \lim_{\vec{r}\to \vec{r}'_{0}\in\partial \Sigma}\Big(\vec{\nabla}G(\vec{r},\vec{r}_{0})\cdot \hat{n}\Big)=\Big(\vec{\nabla}G(\vec{r},\vec{r}_{0})\cdot \hat{n}\Big)\Big|_{\vec{r}'_{0}\in\partial \Sigma}=0
\end{equation}
\\
Finally, a small caveat deserves to be mentioned. Doing a volume integration over \eqref{G0} and using our choice that,$\Big(\vec{\nabla}G(\vec{r},\vec{r}_{0})\cdot \hat{n}\Big)\Big|_{\partial \Sigma}=0$, we get,
\begin{align}
    &\int_{\Sigma}~\vec{\nabla}^2G(\vec{r},\vec{r}_{0})~dV+\int_{\Sigma}~\kappa^2 G(\vec{r},\vec{r}_{0})~dV~=~\int_{\Sigma}\Tilde{\delta}(\vec{r}-\vec{r}_{0})~dV\\
    &\Rightarrow \int_{\partial \Sigma}\Big(\vec{\nabla}G(\vec{r},\vec{r}_{0})\cdot \hat{n}\Big)\Big|_{\partial \Sigma}~dS_{0}+\kappa^2 \int_{\Sigma}G(\vec{r},\vec{r}_{0})~dV=\dfrac{1}{2}\\
    &\Rightarrow\kappa^2 \int_{\Sigma}G(\vec{r},\vec{r}_{0})~dV=\dfrac{1}{2}\label{GFC}
\end{align}
\\
The volume integration vanishes for the homogeneous Neumann boundary condition.
For, $\kappa^2 = \omega^2 - m^2 =0$, the above equation is incorrect. Therefore, this method is only restricted to non-zero values of
$\kappa \neq 0$.\\

\subsection{Solutions of KG Equation in Different Dimensions}
In this section we will solve the Klein-Gordon equation \eqref{EQNM-1} in finite, bounded regions in one, two, and three dimensional spaces. We choose a closed, connected interval $\Big[0,L\Big]$ in one dimension, a disc or radius `$a$', and  a sphere of radius `$a$' in three dimensions, inside which field is confined. As mentioned previously, all these domains are radially symmetric.
\subsubsection{One dimensional interval~$0\leqslant x\leqslant L$}
The one dimensional Green's function corresponding to the Helmholtz equation is,
\begin{equation}
\label{GF1}
    \dfrac{d^2 G(x,x_0)}{dx^2}~+~\kappa^2 ~G(x,x_0)~=~\Tilde{\delta}(x-x_0)
\end{equation}
In order to find $G(x,x_0)$, we apply the eigenfunction expansion method. The eigenfunction equation corresponding to Helmholtz equation is,
\begin{equation}
    \dfrac{d^2 \phi_{n}(x)}{dx^2}~+~\lambda_{n}^2~\phi_{n}(x)~=~0
\end{equation}
where, $\phi_{n}(x)$'s are eigenfunctions corresponding to eigenvalues $\lambda_{n}$.
\\
Then $\phi_{n}(x)$ has the following form under homogeneous Neumann boundary conditions,
\begin{equation}
    \phi_{n}(x)=A_{n}~\cos \lambda_{n}x
\end{equation}
where, $\lambda_{n}=\dfrac{n\pi}{L},~n=0,1,2,3,\cdots$
\\
Being a self-adjoint operator, under the homogeneous Neumann boundary condition, the set of eigenfunctions of the Helmholtz operator,$\Big( \cos\dfrac{n \pi x}{L},n=0,1,2,\cdots\Big)$ form a complete set. Using \eqref{Delta-2} we can expand the delta functions in terms of $\cos \lambda_{n}x$,
\begin{align}
\label{DE1}
    &\Tilde{\delta}(x-0)=\dfrac{1}{2L}+\dfrac{1}{L}\sum_{n=1}^{\infty}\cos{\dfrac{n\pi x}{L}}\\
    &\Tilde{\delta}(x-L)=\dfrac{1}{2L}+\dfrac{1}{L}\sum_{n=1}^{\infty}(\cos{n\pi})~\cos{\dfrac{n\pi x}{L}}\label{DE2}
\end{align}
 The Green's function $G(x,x_{0})$ can also be expanded in the similar way. We have, 
\begin{equation}
\label{GF2}
    G(x,x_{0})~=~A_{0}(x_{0})+\sum_{n=1}^{\infty}A_{n}(x_{0})~\cos\dfrac{n \pi x}{L}
\end{equation}
Putting \eqref{GF2},\eqref{DE1} and \eqref{DE2} into \eqref{GF1}  we get, 
\begin{equation}
\label{GF3}
    G(x,0)=\dfrac{1}{2\kappa^2 L}+\sum_{n=1}^{\infty}\dfrac{1}{L\Big(\kappa^2-\dfrac{n^2\pi^2}{L^2}\Big)}\cos{\dfrac{n \pi x}{L}}
\end{equation}
and
\begin{equation}
\label{GF4}
    G(x,L)=\dfrac{1}{2\kappa^2 L}+\sum_{n=1}^{\infty}\dfrac{\cos{n\pi}}{L\Big(\kappa^2-\dfrac{n^2\pi^2}{L^2}\Big)}\cos{\dfrac{n \pi x}{L}}
\end{equation}
The given boundary conditions are ,
\begin{equation}
\label{BCO}
    \vec{\nabla} \psi\cdot\hat{n}\Big|_{x=0}=-\alpha;~~~\vec{\nabla} \psi\cdot \hat{n}\Big|_{x=L}=\beta
\end{equation}
Here, we choose $\it{negative}$ `$\alpha$' to be consistent with the opposite increment of graphs at two end points in Figure \ref{Weight_Function}.
Therefore, from \eqref{GF3},\eqref{GF4},\eqref{BCO} and \eqref{DI} we get, for the particular integral $\phi_1 ({\vec x})$, defined in \eqref{DI},
\begin{align}
    &\Phi_{1}(\vec{x})~=~G(x,0)~\vec{\nabla} \psi\cdot\hat{n}\Big|_{x=0}~+~G(x,L)~\vec{\nabla} \psi\cdot \hat{n}\Big|_{x=L}\\
    &=-\alpha~\Big[\dfrac{1}{2\kappa^2 L}+\sum_{n=1}^{\infty}\dfrac{1}{L\Big(\kappa^2-\dfrac{n^2\pi^2}{L^2}\Big)}\cos{\dfrac{n \pi x}{L}}\Big]+\beta \Big[\dfrac{1}{2\kappa^2 L}+\sum_{n=1}^{\infty}\dfrac{\cos{n\pi}}{L\Big(\kappa^2-\dfrac{n^2\pi^2}{L^2}\Big)}\cos{\dfrac{n \pi x}{L}}\Big] \label{DIO}
\end{align}
The homogeneous solution $\Phi_{0}(x)$ of the equation \eqref{HOM} is,
\begin{equation}
    \Phi_{0}(x)=A \cos~\kappa x~+~B~\sin~\kappa x
\end{equation}
From \eqref{BC2} and \eqref{BCO} we get,
\begin{equation}
\label{HSO}
    \Phi_{0}(x)~=~\dfrac{\alpha}{\kappa \sin ~\kappa L}~\cos~\kappa(L-x)~-~\dfrac{\beta}{\kappa \sin ~\kappa L}~\cos~\kappa x
\end{equation}

To conclude, the total solution obtained by combining the homogeneous solution \eqref{HSO} and the particular integral \eqref{DIO} is given by 
\begin{equation}
    \psi(\vec{x})=\Phi_{0}(\vec{x})+\Phi_{1}(\vec{x})
\end{equation}
It is straightforward to check that, both $G(x,0)$  and $G(x,L)$ are the Fourier series of $\dfrac{\cos \kappa(L-x)}{2 \kappa \sin \kappa L}$ and $\dfrac{\cos \kappa x}{2 \kappa \sin \kappa L}$ over the intervals $[0,2L]$ and $[-L,L]$ respectively. Following the discussion in the previous section, differentiating each terms of both series and putting the values of end point coordinates, we find that both converge to zero. In other hand, 
\begin{align}
    &\dfrac{d}{dx}\Big[\dfrac{\cos \kappa(L-x)}{2 \kappa \sin \kappa L}\Big]\Big|_{x=0}=\dfrac{1}{2};~~~\dfrac{d}{dx}\Big[\dfrac{\cos \kappa x}{2 \kappa \sin \kappa L}\Big]\Big|_{x=0}=0\label{jump_1}\\
    &\dfrac{d}{dx}\Big[\dfrac{\cos \kappa(L-x)}{2 \kappa \sin \kappa L}\Big]\Big|_{x=L}=0;~~~\dfrac{d}{dx}\Big[\dfrac{\cos \kappa x}{2 \kappa \sin \kappa L}\Big]\Big|_{x=L}=-\dfrac{1}{2}\label{jump_2}
\end{align}
indicates that, derivative of the Green's function, when $x=x_{0}$, has a jump discontinuity \eqref{G1}. Therefore, from \eqref{jump_1},\eqref{jump_2},
\begin{align}
&\lim_{x \to 0}\vec{\nabla}G(x,0)\cdot \hat{n}=\lim_{x \to 0}\dfrac{dG(x,0)}{dx}\hat{i}\cdot (\hat{-i})=-\lim_{x \to 0}\dfrac{dG(x,0)}{dx}=-\dfrac{1}{2}\\
&\lim_{x \to L}\vec{\nabla}G(x,L)\cdot \hat{n}=\lim_{x \to L}\dfrac{dG(x,L)}{dx}\hat{i}\cdot (\hat{i})=\lim_{x \to L}\dfrac{dG(x,L)}{dx}=-\dfrac{1}{2}
\end{align}
we get,
\begin{align}
    \lim_{x \to 0}\vec{\nabla}\Phi_{1}\cdot \hat{n}&=(-\alpha)\lim_{x \to 0}\vec{\nabla}G(x,0)\cdot \hat{n}+\beta\lim_{x \to 0}\vec{\nabla}G(x,L)\cdot \hat{n}\\
    &=\dfrac{\alpha}{2}
\end{align}
and, 
\begin{align}
    \lim_{x \to L}\vec{\nabla}\Phi_{1}\cdot \hat{n}&=(-\alpha)\lim_{x \to L}\vec{\nabla}G(x,0)\cdot \hat{n}+\beta\lim_{x \to L}\vec{\nabla}G(x,L)\cdot \hat{n}\\
    &=-\dfrac{\beta}{2}
\end{align}
The above results satisfy \eqref{limpoi}.
\subsubsection{A Disc in Two Dimensions:}
Let us consider a disc of radius `$a$' on a two dimensional plane, centering the origin. The Green's function corresponding to the Helmholtz equation inside that disc, satisfying the $\it{homogeneous}$ Neumann boundary condition, is given by \cite{Duffy2015},
\begin{equation}
\label{GF21}
    G(\vec{r},\vec{r}_{0})=\dfrac{1}{2\pi a^2 \kappa^2}-\dfrac{1}{\pi}\sum_{n=0}^{\infty}\sum_{m=1}^{\infty}\dfrac{J_{n}(\kappa_{nm}r)J_{n}(\kappa_{nm}r_{0})\cos[n(\theta-\theta_{0})]}{\epsilon_{n}\Big(a^2-\dfrac{n^2}{\kappa_{nm}^2}\Big)\Big(\kappa_{nm}^2-\kappa^2\Big)J^2_{n}(\kappa_{nm}a)}
\end{equation}
Where, $\epsilon_{0}=2, \epsilon_{n}=1$ for $n>0$. $\kappa_{nm}$ are the roots of $\dfrac{d}{dr}~J_{n}(\kappa_{nm}r)\Big|_{r=a}=0$ and $\vec{r}_{0} \in \partial\Sigma.$
\\
From \eqref{G0}, \eqref{Delta-1} and \eqref{Delta-2} the Green's function in \eqref{GF21} is $\dfrac{1}{2}$ of the usual Green's function in a two dimensional disc.
\\
Putting \eqref{GF21} into \eqref{DI} with the Neumann boundary condition $(\vec{\nabla}\psi\cdot \hat{n})\Big|_{\partial\Sigma}=\dfrac{\partial \psi}{\partial r}\Big|_{r=a}=f(\theta_{0})$, we get $\Phi_{1}(r, \theta)$ in terms of Fourier coefficients of $f(\theta_{0})$,
\\
\begin{align}
    \Phi_{1}(r, \theta)&=\int_{\theta_{0}=0}^{2 \pi}\Big[\dfrac{1}{2\pi a^2 \kappa^2}\Big]f(\theta_{0})~a~d\theta_{0}\\
    &-\dfrac{1}{\pi}\sum_{n=0}^{\infty}\sum_{m=1}^{\infty}\dfrac{J_{n}(\kappa_{nm}r)J_{n}(\kappa_{nm}a)}{\epsilon_{n}\Big(a^2-\dfrac{n^2}{\kappa_{nm}^2}\Big)\Big(\kappa_{nm}^2-\kappa^2\Big)J^2_{n}(\kappa_{nm}a)}\int_{\theta=0}^{2 \pi}f(\theta_{0})\cos[n(\theta-\theta_{0})]~a~d\theta_{0}\\
    &=\dfrac{A_{0}}{2a\kappa^2}- \sum_{n=0}^{\infty}\sum_{m=1}^{\infty}\dfrac{J_{n}(\kappa_{nm}r)\Big(A_{n}\cos{n \theta}+B_{n}\sin{n \theta}\Big)}{\epsilon_{n}a\Big(1-\dfrac{n^2}{a^2\kappa_{nm}^2}\Big)\Big(\kappa_{nm}^2-\kappa^2\Big)J_{n}(\kappa_{nm}a)}\label{DIT1}
\end{align}
where, 
\begin{equation}
\label{110}
    f(\theta_{0})=\dfrac{A_{0}}{2}+\sum_{n=1}^{\infty}A_{n}\cos{n \theta_{0}}+\sum_{n=1}^{\infty}B_{n}\sin{n \theta_{0}}
\end{equation}
\\
$f(\theta_{0})$ has been expanded in a Fourier series with period $2\pi$. The Fourier coefficients are the following, 
\\
\begin{align}
    &A_{0}=\dfrac{1}{\pi}\int_{\theta_{0}=0}^{2 \pi}f(\theta_{0})~d\theta_{0}\label{FC21}\\
    &A_{n}=\dfrac{1}{\pi}\int_{\theta_{0}=0}^{2 \pi}f(\theta_{0})\cos{n \theta_{0}}~d\theta_{0}\label{FC22}\\
    &B_{n}=\dfrac{1}{\pi}\int_{\theta_{0}=0}^{2 \pi}f(\theta_{0})\sin n{\theta_{0}}~d\theta_{0} \label{FC23}
\end{align}
The homogeneous solution of the Helmholtz equation inside a disc on a two dimensional plane is the following \cite{Riley2006},
\begin{equation}
    \Phi_{0}(r, \theta)=\sum_{n=0}^{\infty}~C_{n}J_{n}(\kappa r)~\cos{n \theta}+\sum_{n=0}^{\infty}~D_{n}J_{n}(\kappa r)~\sin{n \theta}
\end{equation}
\begin{equation}
    =C_{0}J_{0}(\kappa r)+\sum_{n=1}^{\infty}~C_{n}J_{n}(\kappa r)~\cos{n \theta}+\sum_{n=1}^{\infty}~D_{n}J_{n}(\kappa r)~\sin{n \theta}\label{HST1}
\end{equation}
\\
As the Neumann functions diverge at origin, all coefficients of those functions are zero.
\\
We have the $inhomogeneous$ Neumann boundary condition $\lim_{\vec{r}\to \partial \Sigma}(\vec{\nabla}\Phi_{0} \cdot \hat{n})=\\(\vec{\nabla}\Phi_{0} \cdot \hat{n})\Big|_{\partial \Sigma}=f(\theta_{0})$ as \eqref{BC2}. Then from \eqref{FC21},\eqref{FC22} and \eqref{FC23} we get, 
\\
\begin{align}
\label{HST12}
    \Phi_{0}(r,\theta)=\dfrac{A_{0}}{2 \kappa  J'_{0}(\kappa a)}J_{0}(\kappa r)+\sum_{n=1}^{\infty}\dfrac{A_{n}J_{n}(\kappa r )}{\kappa J'_{n}(\kappa a)}\cos{n \theta}+\sum_{n=1}^{\infty}\dfrac{B_{n}J_{n}(\kappa r )}{\kappa J'_{n}(\kappa a)}\sin{n \theta}
\end{align}
\\
The above solution only exists when $J'_{n}(\kappa a) \neq 0$. Hence, we have to fix `$\kappa$' in Helmholtz equation such that, it is not a root of $J'_{n}(\kappa a) = 0$.
\\
Thus, the solution of the Helmholtz equation inside a disc of radius `$a$' is the sum of \eqref{HST12} and \eqref{DIT1},
\begin{equation}
    \psi(\vec{r})=\Phi_{0}(\vec{r})+\Phi_{1}(\vec{r})
\end{equation}
\subsubsection{Three dimensional sphere of radius $a$:}
As is the standard practice, we use separation of variable to break the Helmholtz equation into three independent equations,
\begin{align*}
    &\dfrac{d^2 \Phi}{d\phi^2}+m^2 \Phi=0\\
    &\dfrac{1}{\sin \theta}\dfrac{d}{d \theta}\Big(\sin \theta \dfrac{d\Theta}{d \theta}\Big)+\Big[n(n+1)-\dfrac{m^2}{\sin^2 \theta}\Big]\Theta=0\\
    &\dfrac{1}{r^2}\dfrac{d}{dr}\Big(r^2\dfrac{dR}{dr}\Big)+\Big[\kappa^2-\dfrac{n(n+1)}{r^2}\Big]R=0
\end{align*}
The Green's function corresponding to the Helmholtz equation inside a sphere of radius $a$, satisfying the homogeneous Neumann boundary condition is given by \cite{houston1954methods},
\begin{align}
\label{GFTH1}
    G(\vec{r}, \vec{r}_{0})~=~\sum_{n=0}^{\infty}\sum_{m=0}^{n}\sum_{s}\dfrac{1}{2\Lambda_{mns}^2\Big(\kappa^2-\dfrac{\pi^2\alpha^2_{ns}}{a^2}\Big)}\Big[\Psi^{e}_{mns}(\vec{r}_{0})\Psi^{e}_{mns}(\vec{r})+\Psi^{o}_{mns}(\vec{r}_{0})\Psi^{o}_{mns}(\vec{r})\Big]
\end{align}
where, 
\begin{align}
    &\Psi^{e}_{mns}(\vec{r})=\cos{m \phi}~ P_{n}^{m}(\cos{\theta})~j_{n}(\dfrac{\pi \alpha_{ns}r}{a})\\
    &\Psi^{o}_{mns}(\vec{r})=\sin{m \phi}~ P_{n}^{m}(\cos{\theta})~j_{n}(\dfrac{\pi \alpha_{ns}r}{a})\\
    &\Lambda_{mns}^2=\dfrac{(\dfrac{2 \pi a^3}{\epsilon_{m}})(n+m)!}{(2n+1)(n-m)!}\Big[\dfrac{(\pi \alpha_{ns})^2-n(n+1)}{(\pi \alpha_{ns})^2}\Big]j_{n}^2(\pi \alpha_{ns})
\end{align}
Here, $\Psi_{mns}^{e}$ and $\Psi_{mns}^{o}$ are mutually orthogonal eigenfunctions of the Helmholtz operator.
$\alpha_{ns}$ are the roots of $\dfrac{d j_{n}(\pi \alpha)}{d\alpha}\Big|_{r=a}=0$, satisfying the homogeneous Neumann boundary condition on the surface of the sphere and $\vec{r}_{0}\in \partial \Sigma$. Here, $\epsilon_{0}=1,\epsilon_{m}=2~~\hbox{for},m=1,2,3,\cdots$. Then from \eqref{DI} we have,
\begin{equation}
\label{DITH1}
    \Phi_{1}(\vec{r})~=~\int_{\partial \Sigma}~G(\vec{r},\vec{r}_{0})(\nabla\psi \cdot \hat{n})|_{\partial \Sigma}~dS_{0}
\end{equation}
We choose the boundary condition as $(\nabla\psi \cdot \hat{n})|_{\partial \Sigma}=f(\theta_{0}, \phi_{0})$. Any function of angular coordinates on a sphere can be written as linear combination of spherical harmonics,\\~$Y_{n}^{m}(\theta, \phi)=\sqrt{\dfrac{2n+1}{4\pi}\dfrac{(n-m)!}{(n+m)!}}~P_{n}^{m}(\cos \theta)~e^{im \phi}$.
\begin{align}
    f(\theta, \phi)&=\sum_{n=0}^{\infty}\sum_{m=-n}^{n}A_{mn}~Y_{n}^{m}(\theta, \phi)\\
    &=\sum_{n=0}^{\infty}C_{n,0}P_{n}^{0}(\cos \theta)+\sum_{n=0}^{\infty}\sum_{m=1}^{n}C_{n,m}P_{n}^{m}(\cos \theta)\cos{m \phi}\\
    &+\sum_{n=0}^{\infty}\sum_{m=1}^{n}D_{n,m}P_{n}^{m}(\cos \theta)\sin{m \phi}
\end{align}
Using the orthogonality property of associated Legendre polynomials,
\begin{align}
    &\int_{\theta=0}^{\pi}d\theta~\sin\theta~P_{n}^{m}(\cos\theta)~P_{n'}^{m}(\cos\theta)~=~\dfrac{2}{2n+1}\dfrac{(n+m)!}{(n-m)!}\delta_{nn'}
\end{align}
we get, 
\begin{align}
    &C_{n,0}=\dfrac{2n+1}{4 \pi}\int_{\theta_{0}=0}^{\pi}\int_{\phi_{0}=0}^{2 \pi}f(\theta_{0}, \phi_{0})P_{n}^{0}(\cos \theta_{0})\sin \theta_{0} ~d\theta_{0} ~d\phi_{0}\\
    &C_{n,m}=\dfrac{(2n+1)(n-m)!}{2 \pi (n+m)!}\int f(\theta_{0},\phi_{0})P_{n}^{m}(\cos{\theta_{0}})\cos{m\phi_{0}} \sin{\theta_{0}}~d\theta_{0} ~d\phi_{0}\\
    &D_{n,m}=\dfrac{(2n+1)(n-m)!}{2 \pi (n+m)!}\int f(\theta_{0},\phi_{0})P_{n}^{m}(\cos{\theta_{0}})\sin{m\phi_{0}} \sin{\theta_{0}}~d\theta_{0}~ d\phi_{0}
\end{align}
Putting \eqref{GFTH1} into \eqref{DITH1} we get,\Big(here, $\kappa_{ns}=\dfrac{\pi \alpha_{ns}}{a}$\Big)
\begin{align}
    \Phi_{1}(\vec{r})=\sum_{n=0}^{\infty}\sum_{m=0}^{n}\sum_{s}\dfrac{1}{2\Lambda_{mns}^2(\kappa^2-\kappa_{ns}^2)}&\Big[\int \Psi_{mns}^{e}(\vec{r})
     \Psi_{mns}^{e}(\vec{r}_{0})f(\theta_{0}, \phi_{0})~dS_{0}\\
     &+\int \Psi_{mns}^{o}(\vec{r})
     \Psi_{mns}^{o}(\vec{r}_{0})f(\theta_{0}, \phi_{0})~dS_{0}\Big]
\end{align}
\begin{align}
     &=\sum_{n=0}^{\infty}\sum_{s}\dfrac{1}{2\Lambda_{0ns}^2(\kappa^2-\kappa_{ns}^2)}\Psi_{0ns}^{e}(\vec{r})\times \dfrac{4 \pi a^2}{2n+1} C_{n,0}~j_{n}(\pi \alpha_{ns})\\
     &+\sum_{n=0}^{\infty}\sum_{m=1}^{n}\sum_{s}\dfrac{1}{2\Lambda_{mns}^2(\kappa^2-\kappa_{ns}^2)}\Psi_{mns}^{e}(\vec{r})\times \dfrac{2 \pi a^2 (n+m)!}{(2n+1)(n-m)!} C_{n,m}~j_{n}(\pi \alpha_{ns})\\
     &+\sum_{n=0}^{\infty}\sum_{m=1}^{n}\sum_{s}\dfrac{1}{2\Lambda_{mns}^2(\kappa^2-\kappa_{ns}^2)}\Psi_{mns}^{o}(\vec{r})\times \dfrac{2 \pi a^2(n+m)!}{(2n+1)(n-m)!} D_{n,m}~j_{n}(\pi \alpha_{ns})\label{DITH2}
\end{align}
\\
\\
The homogeneous solution of the Helmholtz equation inside a sphere in three-dimensional spherical polar coordinates is, 
\begin{equation}
\label{HSTH}
    \Phi_{0}(\vec{r})~=~\sum_{n=0}^{\infty}\sum_{m=0}^{n}~E_{nm}j_{n}(\kappa r) Y_{n}^{m}(\theta, \phi)
\end{equation}
Applying, \eqref{BC2} We get, 
\begin{equation}
   f(\theta_{0},\phi_{0})=\sum_{n=0}^{\infty}\sum_{m=0}^{n}E_{nm}\kappa j'_{n}(\kappa a)Y_{n}^{m}(\theta_{0}, \phi_{0})
\end{equation}
Using the orthonormality property of spherical harmonics,
\begin{equation}
    \int_{\theta=0}^{\pi}\int_{\phi=0}^{2\pi}~d\theta~\sin\theta~d\phi~Y_{n}^{m}Y_{n'}^{m'*}=\delta_{nn'}\delta_{mm'}
\end{equation}
we get, 
\begin{equation}
\label{COFF}
    E_{nm}=\dfrac{1}{\kappa j'_{n}(\kappa a)}\int_{\theta_{0}=0}^{\pi}\int_{\phi_{0}=0}^{2 \pi} f(\theta_{0}, \phi_{0})~Y_{n}^{m*}(\theta_{0}, \phi_{0})~\sin{\theta_{0}}~d \theta_{0} ~d \phi_{0}
\end{equation}
Like in one and two dimensions, in the case of a sphere, the coefficients $E_{nm}$ do not allow the values of $\kappa$, which are the roots of $j'_{n}(\kappa a)=0$. Therefore, the total solution for the sphere from \eqref{HSTH} with \eqref{COFF} and \eqref{DITH2},
\begin{equation}
    \psi(\vec{r})=\Phi_{0}(\vec{r})+\Phi_{1}(\vec{r})
\end{equation}
\section{Conclusion and Discussion:}
In this paper we have used the formalism  of Bering to study the Klein-Gordon theory on a space with a boundary. The presence of the boundary modifies the equation in a non-trivial fashion. The modified equation is solved in various dimensions under the assumption of appropriate symmetry. The solutions thus obtained encode the information about the boundary.\\ 

It should be added that the scattering of classical fields from obstructions is a 
well-studied branch  of applied mathematics. In the present paper, we
studied the theory inside a bounded domain, with the boundary acting like an insurmountable topological obstruction.
The non-trivial correction due to the presence of the boundary along with the the condition of impenetrability is 
realised through the use of {\it inhomogeneous} Neumann boundary conditions on the modified Klein-Gordon equation.\\

\section{Acknowledgement} This work is partially funded by a grant from the Infosys Foundation. We thank to Prof. Ghanashyam Date for valuable discussions.\\
\section{Appendix }
In this section we have put the notations used throughout the paper. However, these are mostly related to the review to the Bering's paper. One can go through that paper \cite{Bering:1998fm} and also \cite{Soloviev:1993uxk} for more details.\\
\subsection{Notations:}
1.
\begin{equation}
    \partial^k={\partial_{1}}^{k_{1}}{\partial_{2}}^{k_{2}}\cdots {\partial_{d}}^{k_{d}}
\end{equation}
\begin{equation}
    =\left(\frac{\partial}{\partial x_{1}}\right)^{k_{1}}\left(\frac{\partial}{\partial x_{2}}\right)^{k_{2}}\cdots\left(\frac{\partial}{\partial x_{d}}\right)^{k_{d}}  
\end{equation}
\\
(a).Where, $k_{i}$'s take any values from $0$ to $\infty$.But all $k_{i}$'s will not be zero simultaneously.e.g,
\begin{equation}
    (k_{1},k_{2},\cdots ,k_{d})\in {N_{0}}^{d}/\{(0,0,\cdots ,0)\}, N_{0}=\{0,1,2,3,\cdots\}
\end{equation}
(b).`$d$' is the dimension of the space.
\\
\\
2.The higher partial derivative is defined as following,
\begin{equation}
\label{174}
    P_{A(k)}f(x)=\dfrac{\partial f(x)}{\partial \phi^{A(k)}(x)}=\dfrac{\partial f(x)}{\partial\left(\dfrac{\partial^k \phi^{A}(x)}{\partial x_{1}^{k_{1}} \partial x_{2}^{k_{2}} \dots \partial  x_{d}^{k_{d}}}\right)}
\end{equation}
where,
\begin{equation}
\label{175}
    k=k_{1}+k_{2}+\cdots +k_{d}, ~f=f(\partial^k \phi^{A}(x),x)
\end{equation}
$\phi^{A}(x,t)$ are the classical fields depending on space and time. $A=1,2,3,\cdots$ are the number of fields involved in the system. We are interested in spatial derivatives, where time will not be involved. Throughout the entire formalism we have maintain that.
\\
\\
3.We have the following notation,
\begin{equation}
    \dfrac{\partial f(x)}{\partial\phi^{A(k=0)}(x)}=\dfrac{\partial f(x)}{\partial \phi^{A}(x)}
\end{equation}
\\
4.~A \textit{Functional} in a local form is defined as following,
\begin{equation}
    F[\phi^{A}]=\int d^d x f(\partial^k \phi^{A}(x),x)
\end{equation}
We have two formulation of $\delta F$, which are expressed in terms of higher partial derivative and higher functional derivative.
\begin{align}
    \delta F &=\int_{\sum}d^d x
              \sum_{k=0}^{\infty} \partial^{k}\Big[ \dfrac{\delta F}{\delta \phi^{A(k)}(x)}\delta \phi^{A}(x)\Big]\\
             &=\int_{\sum}d^d x \sum_{k=0}^{\infty}\dfrac{\partial F}{\partial \phi ^{A(k)}(x)} \partial ^{k}\left(\delta \phi^{A}(x)\right) 
\end{align}
Where, higher partial and functional derivatives are connected by the following expression,
\begin{equation}
\dfrac{\delta F}{\delta \phi^{A(k)}(x)}=\sum_{m \ge k} \binom{m}{k}(-\partial)^{m-k}\dfrac{\partial F}{\partial \phi ^{A(m)}(x)}
\end{equation}
In terms of local form we choose them as,
\begin{equation}
    \dfrac{\partial F}{\partial \phi^{A(k)}(x)}=P_{A(k)}f(x)=\dfrac{\partial f(x)}{\partial \phi^{A(k)}(x)}
\end{equation}
And,
\begin{equation}
    \dfrac{\delta F}{\delta \phi^{A(k)}(x)}=\sum_{m \ge k} \binom{m}{k}(-\partial)^{m-k}\dfrac{\partial f(x)}{\partial \phi^{A(m)}(x)}=\sum_{m \ge k} \binom{m}{k}(-\partial)^{m-k}P_{A(m)}f(x)=E_{A(k)}f(x)
\end{equation}
\subsection{Rules for Higher Euler Operator $E_{A(k)}$:}\label{7.2}
Here, we have discussed how to use Euler Operator $E_{A(k)}$, to derive the equation of motion from \eqref{Field} to \eqref{Momentum}.
\begin{equation}
E_{A(k)}f(x)=\sum_{m \ge k} \binom{m}{k}(-\partial)^{m-k}\dfrac{\partial f(x)}{\partial \phi^{A(m)}(x)}
\end{equation}
The Binomial coefficients for multi index,
\begin{equation}
    \binom{m}{k}=\binom{m_1}{k_1}\binom{m_2}{k_2}\cdots\binom{m_d}{k_d}
\end{equation}
Where, the ordinary binomial coefficients are, 
\begin{align}
    \binom{m_i}{k_i}&=\dfrac{m_{i}!}{k_{i}!(m_{i}-k_{i})!}~~\hbox{when},~0 \leq k_{i} \leq m_{i}\\
    &=0~~~~~~~~~~~~~~~~~~\hbox{otherwise}
\end{align}
and, 
\begin{equation}
    (-\partial)^{m-k}=(-\partial_{1})^{m_{1}-k_{1}}(-\partial_{2})^{m_{2}-k_{2}}\cdots(-\partial_{d})^{m_{d}-k_{d}}
\end{equation}
For, example in \eqref{43},
\begin{multline}
   \sum_{m \geq 0} \binom{m}{0} (-\partial)^{m}\Big(\dfrac{\partial (\chi_{\epsilon}\mathscr{H})}{\partial \phi^{  (m)}}\Big)\\
  =\binom{0}{0}\binom{0}{0}\binom{0}{0}(-\partial)_{x}^{0}
  (-\partial)_{y}^{0}(-\partial)_{z}^{0}\Bigg(\dfrac{\partial (\chi_{\epsilon}\mathscr{H})}{\partial \phi^{(m=0)}}\Bigg)+\binom{1}{0}\binom{0}{0}\binom{0}{0}(-\partial)_{x}^{1}
  (-\partial)_{y}^{0}(-\partial)_{z}^{0}\Bigg(\dfrac{\partial (\chi_{\epsilon}\mathscr{H})}{\partial\Big(\dfrac{\partial \phi}{\partial x^{1}\partial y^{0}\partial z^{0}}\Big)}\Bigg)+\\
  \binom{0}{0}\binom{1}{0}\binom{0}{0}(-\partial)_{x}^{0}
  (-\partial)_{y}^{1}(-\partial)_{z}^{0}\Bigg(\dfrac{\partial (\chi_{\epsilon}\mathscr{H})}{\partial \Big(\dfrac{\partial \phi}{\partial x^{0}\partial y^{1}\partial z^{0}}\Big)}\Bigg)+\binom{0}{0}\binom{0}{0}\binom{1}{0}(-\partial)_{x}^{0}
  (-\partial)_{y}^{0}(-\partial)_{z}^{1}\Bigg(\dfrac{\partial (\chi_{\epsilon}\mathscr{H})}{\partial \Big(\dfrac{\partial \phi}{\partial x^{0}\partial y^{0}\partial z^{1}}\Big)}\Bigg)\label{188}
\end{multline}
\begin{equation}
=\Big[\chi_{\epsilon}\dfrac{\partial \mathscr{H}}{\partial \phi}-\dfrac{\partial}{\partial x}\Big(\dfrac{\partial(\chi_{\epsilon} \mathscr{H})}{\partial \Big(\dfrac{\partial \phi}{\partial x}\Big)}\Big)-\dfrac{\partial}{\partial y}\Big(\dfrac{\partial(\chi_{\epsilon} \mathscr{H})}{\partial \Big(\dfrac{\partial \phi}{\partial y}\Big)}\Big)-\dfrac{\partial}{\partial z}\Big(\dfrac{\partial(\chi_{\epsilon} \mathscr{H})}{\partial \Big(\dfrac{\partial \phi}{\partial z}\Big)}\Big)\Big]
\end{equation}
Where, the first term in \eqref{188} is corresponding to $m=0$, that has only one possibility, $m_{1}=0,m_{2}=0,m_{3}=0$, and among last three terms, each one is corresponding to $m=1$, having three possibilities ($m_{1}=1,m_{2}=0,m_{3}=0$),($m_{1}=0,m_{2}=1,m_{3}=0$)and ($m_{1}=0,m_{2}=0,m_{3}=1$). Here, in $m-k$, $k$ is always zero, e.g., $k_{1}=0,k_{2}=0,k_{3}=0$.\\
Further details about higher Euler operators can be found in \cite{Soloviev:1993uxk} and \cite{Olver1986}.
\subsection{Abstract Manifold}\label{7.3}
We can generalize the idea into an abstract manifold, where the notations and few relations have been modified in order to find the equation of motion. The construction will be generalized in a chart $\sum \subseteq R^d$. The derivative will be modified from spatial derivative to covariant derivative $D_{i}=\partial_{i}+\Gamma_{i}$ and the trivial measure $d^d x$ will be replaced by $\rho(x) d^d x$. We will assume that, both `$\rho$' and `$D_{i}'$ will not depend on dynamical fields $\phi^A (x,t)$ nor on time $t$ and $D_{i}\rho$=0. The main difference between spatial derivative and covariant derivative is that, where spatial derivatives do commute with each other, covariant derivatives do not. Here, one important point to be noted, spatial derivative also does not commute with higher partial derivative $P_{A(k)}$.
\begin{equation}
\label{PD}
    \partial_{i}=\phi^{A(k+e_i)}P_{A(k)}+\partial_{i}^{explicit}
\end{equation}
Where, $\partial_{i}$ is the total derivative and $\partial_{i}^{explicit}$ is the explicit derivative with respect to $i$ th coordinate. $e_{i}$ is an $n$ tuple whose $i$ th element is 1, others are all zero.
\begin{equation}
    e_{i}=(0,0,\cdots,1,\cdots,0)
\end{equation}
In case of covariant derivative, \eqref{PD} will have the following form,
\begin{equation}
    D_{i}=\phi^{A(i+K)}P_{A(K)}+D_{i}^{explicit}
\end{equation}
and,
\begin{equation}
    \Big[D_{i},D_{j}\Big]=\Big[D_{i}^{explicit},D_{j}^{explicit}\Big]
\end{equation}
Here, $\phi^{A(K)}(x)=D^{K} \phi^{A}(x)$ and 
\begin{equation}
    K=(k_{1},k_2,\cdots,k_{|K|}) \in \{1,\cdots,d\} \times \cdots \times \{1,\cdots,d\}
\end{equation}
Where, `$d$' is the dimension of the space. Although the format is identical but the notation will be modified such that $\binom{n}{m}$ terms in the `$\partial$' notation will be suppressed by a related formalism \cite{Bering:1998fm}. In covariant form, the definition of the functional will have the following form, 
\begin{align}
    \delta F &= \int_{\sum} \rho(x) d^d x \sum_{k=\emptyset}^{\infty} \frac{\partial F}{\partial \phi^{A(K)}(x)} D^{K} \Big(\delta \phi ^{A}(x)\big)\\
            &=\int_{\sum} \rho(x) d^d x \sum_{k=\emptyset}^{\infty} D^{K} \Big[ \frac{\delta F}{\delta \phi^{A(K)}(x)}  \delta \phi ^{A}(x)\Big]\\
 \frac{\delta F}{\delta \phi^{A(K)}(x)}&=\sum_{M \succeq K}(-D)^{M^t \div K^t} \dfrac{\partial F}{\partial \phi^{A(M)}(x)} \\  
 \dfrac{\partial F}{\partial \phi^{A(M)}(x)}   &=\sum_{M \succeq K} D^{M \div K} \dfrac{\delta F}{\delta \phi^{A(M)}(x)} 
\end{align}
We already presented the proof of equation of motion for a classical field in terms of covariant derivative `$D$' in \eqref{Proof of eqnm}.
\subsection{Embedded Approach:}
We will always consider that, our region of interest $\sum$ is always covered by a \textit{single chart} and it is inside a \textit{bounded} region.
The geometric information about the space is encoded inside $\rho$ and $D_{i}$. For example, if $\sum=\textbf{R}^d$, then we need a non trivial $\rho$ and $D_{i}$, such that the entire $\textbf{R}^d$ can be mapped into a bounded region $\tilde{\sum}$ and inside a single chart. Therefore, the spatial boundary infinity is now truly a boundary of $\tilde{\sum}$.
\subsection{Dirac delta distribution:}
Here, we will discuss about two quantities, which have been used through out the entire formalism.
\begin{equation}
    \int_{\Sigma \times \Sigma} \rho(x) d^d x \rho(z) d^d z~~\eta(x,z) \delta_{\Sigma}(x,z)=\int_{\Sigma} \rho(x) d^d x ~\eta(x,x)
\end{equation}
Where,
\begin{equation}
    \delta_{\Sigma}(x,z)=\dfrac{\delta(x-z)}{\chi_{\epsilon}(x)\rho(x)}
\end{equation}
And, \\
we define $\textit{adjoint}$ of the covariant derivative as following, 
\begin{equation}
    D^{\dagger}_{i}f=-\dfrac{1}{\chi_{\epsilon}(x)\rho(x)}D_{i}\left(\chi_{\epsilon}(x)\rho(x)f\right);~\left(-D^{\dagger}_{(x)}\right)^{N}f(x)=\dfrac{1}{\chi_{\epsilon}(x)\rho(x)}D_{(x)}^{N}\left(\chi_{\epsilon}(x)\rho(x)~f(x)\right)
\end{equation}
\subsection{Space of functionals}
In Poisson Bracket \eqref{Poisson Bracket} we used functionals whose coordinates are from different manifolds. Therefore, it is convenient to write down previous expressions in more than one manifolds. We define a functional,
\begin{equation}
\label{functional}
    F(z)=\int_{\sum \times \sum\times \cdots \times \sum} \rho d^dz_{(1)} \cdots \rho d^d z_{(s)}~f(z) 
\end{equation}
Where, 
\begin{equation}
    f(z)=f\left(D^{K_{(1)}}\phi(z_{(1)}),z_{(1)},\cdots,D^{K_{(r)}}\phi(z_{(r)}),z_{(r)}\right)
\end{equation}
Where, $z_{(i)}$ is the space coordinates of $\sum_{(i)}$ and $z_{(i)}\equiv(z_{(i)}^{1},\cdots,z_{(i)}^{d_{i}})$. $d_{i}$ is the dimension of $\sum_{(i)}$.
\\
In left hand side of \eqref{functional}, only those $z_{k}$'s will be inserted which are not integrated out.
\\
We define higher partial derivatives as following,
\begin{equation}
    \dfrac{\partial F}{\partial \phi^{A(K)}(z)}=\int_{\sum^s}
    \rho d^dz_{(1)} \cdots \rho d^d z_{(s)}~\sum_{i=1}^{r}
\delta_{\Sigma}(z_{(i)},x)~P_{A(k)}^{(z_{(i)})} f(z)
\end{equation}
and,
\begin{align}
    \dfrac{\delta F(z)}{\delta \phi^{A(K)}(x)}&=~E_{A(K)}^{(x)} \sum_{i=1}^{r} [ \delta_{\Sigma}(x,z_{(i)})F(z_{(1)},\cdots,z_{(i-1)},x,z_{(i+1)},\cdots)]\\
    &=\sum_{i=1}^{r} \sum_{M \succeq K}\left((-D_{(x)})^{M^t \div K^t} \delta_{\Sigma}(x,z_{(i)})\right)P_{A(M)}^{(z_{(i)})}F(z)
\end{align}

\bibliographystyle{plain}
\bibliography{reference}

\begin{thebibliography}{10}

\bibitem{Aharonov1959}
Y.~Aharonov and D.~Bohm.
\newblock Significance of electromagnetic potentials in the quantum theory.
\newblock {\em Physical Review}, 115(3):485–491, August 1959.

\bibitem{Bering:1998fm}
K.~Bering.
\newblock {Putting an edge to the Poisson bracket}.
\newblock {\em J. Math. Phys.}, 41:7468--7500, 2000.

\bibitem{Berry1980}
M~V Berry.
\newblock Exact aharonov-bohm wavefunction obtained by applying dirac’s magnetic phase factor.
\newblock {\em European Journal of Physics}, 1(4):240–244, October 1980.

\bibitem{Duffy2015}
Dean~G. Duffy.
\newblock {\em Green’s Functions with Applications}.
\newblock Chapman and Hall/CRC, March 2015.

\bibitem{Ehrenberg1949}
W~Ehrenberg and R~E Siday.
\newblock The refractive index in electron optics and the principles of dynamics.
\newblock {\em Proceedings of the Physical Society. Section B}, 62(1):8–21, January 1949.

\bibitem{Gerry1979}
Christopher~C. Gerry and Vijay~A. Singh.
\newblock Feynman path-integral approach to the aharonov-bohm effect.
\newblock {\em Physical Review D}, 20(10):2550–2554, November 1979.

\bibitem{Greiner1996}
Walter Greiner and Joachim Reinhardt.
\newblock {\em Field Quantization}.
\newblock Springer Berlin Heidelberg, 1996.

\bibitem{houston1954methods}
WV~Houston.
\newblock Methods of theoretical physics, pts. i and ii. philip m. morse and herman feshbach. mcgraw-hill, new york-london, 1953. xxx+ 1978 pp. illus. 30theset, 15 each volume.
\newblock {\em Science}, 120(3110):212--213, 1954.

\bibitem{sep-paradox-zeno}
Nick Huggett.
\newblock {Zeno’s Paradoxes}.
\newblock In Edward~N. Zalta and Uri Nodelman, editors, {\em The {Stanford} Encyclopedia of Philosophy}. Metaphysics Research Lab, Stanford University, {S}pring 2024 edition, 2024.

\bibitem{Olver1986}
Peter~J. Olver.
\newblock {\em Applications of Lie Groups to Differential Equations}.
\newblock Springer New York, 1986.

\bibitem{olver2014introduction}
Peter~J Olver et~al.
\newblock {\em Introduction to partial differential equations}, volume~1.
\newblock Springer, 2014.

\bibitem{Riley2006}
K.~F. Riley, M.~P. Hobson, and S.~J. Bence.
\newblock {\em Mathematical Methods for Physics and Engineering}.
\newblock Cambridge University Press, 2006.

\bibitem{Soloviev:1993uxk}
Vladimir~O. Soloviev.
\newblock {Boundary values as Hamiltonian variables. 1. New Poisson brackets}.
\newblock {\em J. Math. Phys.}, 34:5747--5769, 1993.

\end{thebibliography}

\end{document}